# On bi-Hamiltonian structure of some superintegrable systems

Gh.Haghighatdoost[1][*], S. Abdolhadi-Zangakani[2][†]

[1,2]*Department of Mathematics, Azarbaijan Shahid Madani University, 53714-161, Tabriz, Iran*

[2]*Department of Mathematics, University of Bonab, Tabriz, Iran*

December 26, 2017

## Abstract

We discuss bi-Hamiltonian structures for integrable and superintegrable Hamiltonian system on the list of symplectic four-dimensional real Lie groups are classified by G.Ovando in [6]. In addition, we create corresponding control matrix for obtained bi-Hamiltonian structures.

## 1   Introduction

The pioneering work in the bi-Hamiltonian systems is done by Franco Magri [1], Yvette Kosmann-Schwarzbach; Franco Magri [2] and then followed by the fundamental papers by Gelfand and Dorfman[3], Magri and Morosi[4]. These works show that integrabiltiy of many systems in mathematical physics, mechanics, and geometry is closely related to their bi-Hamiltonian structures. It is shown that all classical systems have the bi-Hamiltonian structure, at the same time by using the bi-Hamiltonian methods many new nontrivial and interesting examples of integrable systems have been found. As we know, the study of bi-Hamiltonian structure based on a very simple notion of compatible Poisson structures. It is proved that bi-Hamiltonian structure is very powerful in the theory of integrable Hamiltonian systems not only for finding new examples, but also for integration of systems, constructing separation of variables and description of properties of solutions.

In [5] the integrable Hamiltonian systems with the symmetry Lie group as a four-dimensional phase space where have symplectic structure is constructed. The list of symplectic four dimensional real Lie groups is classified in [6]. Here we find bi-Hamiltonian structures for those integrable Hamiltonian and bi-Hamiltonian systems.

## 2   Preliminaries and the method of calculation

A bi-Hamiltonian manifold M is a smooth manifold endowed with two compatible bi-vectors P, P' such that

$$[P, P] = 0 \ , \ [P, P'] = 0 \ , \ [P', P'] = 0, \qquad (1)$$

where [ , ] is the Schouten bracket.
The bi-vectors P, $P'$ determine a pair of compatible Poisson brackets on M,

$$\{f(x), g(x)\} = < df, Pdg > = \sum_{i,j}^{dimM} P^{i,j}(z) \frac{\partial f(x)}{\partial x_i} \frac{\partial g(x)}{\partial x_j} \qquad (2)$$

for all $f, g \in F(M)$ and similar brackets $\{,\}'$ to $P'$.
Let $H_0, H_1, ..., H_n$ be functionally independent functions on M and in involution with respect to both Poisson brackets

---

[*]Corresponding author, e-mail: gorbanali@azaruniv.ac.ir
[†]e-mail:S.Abdolhadi@bonabu.ac.ir

$$\{H_i, H_j\} = \{H_i, H_j\}' = 0 \qquad i = 0, ..., n , \; j = 0, ..., n. \tag{3}$$

According to [ [7], [8], [9]], let us suppose that the desired second Poisson bivector $P'$ is the Lie derivative of P along some unknown Liouville vector field X

$$P' = L_X(P), \tag{4}$$

which has to satisfy following the equation

$$[P', P'] = [L_X(P), L_X(P)] = 0 \Leftrightarrow [L_X^2(P), P] = 0, \tag{5}$$

with respect to the Schouten bracket [., .]. By definition (4) bivector P is compatible with a given bivector P, i.e. [P, P] = 0.
From all solutions X of the equation (5) we have choosen some solutions X such that

$$\{H_i, H_j\}' = 0 \qquad i = 0, ..., n , \; j = 0, ..., n . \tag{6}$$

Obviously enough, in their full generality the system of equations (4-6) is too difficult to be solved because it has infinitely many solutions labeled by different separated coordinates, see [8] and [10]. In order to get particular solutions we will use some special ansatze for the Liouville vector field X.
According to [11], the bi-involutivity of the integrals of motion (3) is equivalent to the existence of control matrix $F = (F_{ij})$ defined by

$$P' dH_i = P \sum_{j=1,2}^{2} F_{ij} dH_j \qquad i = 1, 2 . \tag{7}$$

When, this matrix be non-degenerate, its eigenvalues are desired variables of seperation ([7], [8]).
In order to solve equations (4-6) we will use polynomials of momenta ansatze for the components of the Liouville vector field $X = \sum X^i \partial_i$

$$X^i = \sum_{k=0}^{N} \sum_{m=0}^{k} g_{km}^i(x_1, x_2) p_1^{k-m} p_2^m . \tag{8}$$

First, we assume $N = 2$, it means that $X^i$ will be generic second order polynomials in momenta $p_1, p_2$ with coefficients $g_{km}^i(x_1, x_2)$ depending on variables $x_1$ and $x_2$. Substituting this ansatz (8) into the equations (4-6) and demanding that all the coefficients at powers of $p_1$ and $p_2$ vanish one gets the over determined system of differential equations which can be solved in the modern computer algebra systems.

In this section, we will construct bi-Hamiltonian structure related to some integrable and super integrable Hamiltonian systems with the real four dimensional real Lie group as a four dimensional phase space [5]. For this propose, we consider those four dimensional real Lie groups such that they have symplectic structure. The list of symplectic four dimensional real Lie groups are classified in [6].

## 3 Construction of bi-Hanmiltonian structures for some four dimensional real Lie groups

Here, we construct the models on the list of symplectic four dimensional real Lie groups are classified by Ovando. We do this separately as follows .
**Lie group $A_{4,1}$:**
According to [5], non-degenerate Poisson structure $P$ on this Lie group can be obtained in the following forms:

$$\{x_1, x_2\} = -c/2 p_2^2, \quad \{x_1, p_1\} = cp_2, \quad \{x_1, p_2\} = -d, \quad \{x_2, p_1\} = -c, \tag{9}$$

Where c and d are arbitrary real constants and $p_1, p_2$ are the conjugate momentum of $x_1$ and $x_2$ respectivlly . The possion structure P can be represented in matrix form as follows:

$$P = \begin{bmatrix} 0 & -c/2p_2^2 & cp_2 & -d \\ c/2p_2^2 & 0 & -c & 0 \\ -cp_2 & c & 0 & 0 \\ d & 0 & 0 & 0 \end{bmatrix}. \tag{10}$$

In this Lie group, we have the superintegrable system with the Hamiltonian H and integrals of motion $H_1, H_2$ as follows[5]:

$$H = -x_2 + \frac{2p_1 p_2}{cd} + \frac{p_2^3}{d^2} + \frac{cp_2^3}{4d} , \tag{11}$$

$$H_1 = (x_1 - \frac{p_1^2}{c^2} - \frac{1}{4}cdx_2 p_2 + \frac{p_1 p_2^2}{4} - \frac{p_1 p_2^2}{cd} + \frac{3c^2 p_2^4}{64} - \frac{p_2^4}{4d^2} + \frac{cp_2^4}{8d})(x_2 - \frac{8dp_1 p_2 + (4c + c^2 d)p_2^3}{4cd^2}) , \tag{12}$$

$$H_2 = \frac{-1}{2}(-x_1 + \frac{p_1^2}{c^2} + \frac{1}{4}cdx_2 p_2 - \frac{p_1 p_2^2}{4} + \frac{p_1 p_2^2}{cd} - \frac{3c^2 p_2^4}{64} + \frac{p_2^4}{4d^2} - \frac{cp_2^4}{8d})^2(x_2 - \frac{8dp_1 p_2 + (4c + c^2 d)p_2^3}{4cd^2}) . \tag{13}$$

The aim is to finding the bi-Hamiltonian structures for given the superintegrable system with integrals of motion $H, H_1, H_2$ on $A_{4,1}$ with Poisson P.

Now by the ansatze (8) and solving the related differential equations, the Poisson Second bi-vector $P'$ will be obtain as follows:

$$P' = \begin{bmatrix} 0 & \alpha & a_4 c + a_3 d + (-a_1 c + a_2 c + 2a_5 d)p_2 & a_2 d \\ * & 0 & a_1 c & 0 \\ * & * & 0 & 0 \\ * & * & * & 0 \end{bmatrix} , \tag{14}$$

where $a_i$s are arbitrary real constants and

$$\alpha = \frac{1}{4cd}(-8d(a_1 - a_2)p_1 + 8(a_4 c + a_3 d)p_2 - (12a_1 c + 20a_2 c + 16a_5 d - a_1 c^2 d + 3a_2 c^2 d)p_2^2) .$$

If the components $X^i$ of the vector field $X$ will be third order polynomials in momenta $p_1, p_2$, then the second Poisson bi-vector will be as follows:

$$P' = \begin{bmatrix} 0 & \beta & a_4 c + a_3 d - (a_1 c + a_2 c + 2a_5 d)p_2 - \frac{1}{2}a_6 c^2 d p_2^2 & a_2 d \\ * & 0 & a_1 c & 0 \\ * & * & 0 & 0 \\ * & * & * & 0 \end{bmatrix} , \tag{15}$$

where $a_i$s are arbitrary real constants and

$$\beta = \frac{2(a_2 - a_1)}{c}p_1 + \frac{2(a_1 d + a_2 c)}{cd}p_2 - a_6 c p_2^3$$
$$+ \frac{16a_5 d - a_1 c^2 d + 3a_2 c^2 d - 12a_1 c + 20a_2 c}{4cd}p_2^2 .$$

In this case, control matrix for the first Poisson bivector $P'$ is obtained as follows:

$$F = \begin{bmatrix} -a_1 & 0 \\ -\frac{\gamma}{64c^2 d^2} & -a_2 \end{bmatrix} , \tag{16}$$

with

$$\gamma = (-16(a_2 c + d(a_3 + 2a_5 p_2)))(8dp_1 p_2 + 4cp_2^3 + c^2 dp_2^3 - 4cd^2 x_2)$$
$$+ a_1(48c^2 p_2^4 + 8d(8cp_1 p_2^2 + 5c^3 p_2^4) - 32c^3 d^3 p_2 x_2 + d^2(-64p_1^2 + 48c^2 p_1 p_2^2$$
$$+ c^2(7c^2 p_2^4 + 64x_1 - 64p_2 x_2))) + a_4(-112c^2 p_2^4 - 8d(24cp_1 p_2^2 + 7c^3 p_2^4) + 32c^3 d^3 p_2 x_2$$
$$+ d^2(64p_1^2 - 48c^2 p_1 p_2^2 + c^2(-7c^2 p_2^4 - 64x_1 + 128p_2 x_2)))) .$$

**Lie group $A_{4,2}^{-1}$:**

The non-degenerate Poisson structure on $A_{4,2}^{-1}$ can be obtained as follows [5]:

$$\{x_1, x_2\} = 2c, \quad \{x_1, p_1\} = -c, \quad \{x_2, p_2\} = -d \ e^{-p_2} , \tag{17}$$

where c and d are arbitrary real constants. and can be represented in matrix form as follows:

$$P = \begin{bmatrix} 0 & 2c & -c & 0 \\ -2c & 0 & 0 & d\, e^{-p_2} \\ c & 0 & 0 & 0 \\ 0 & -d\, e^{-p_2} & 0 & 0 \end{bmatrix}. \tag{18}$$

On this Lie group, the Hamiltonian H and integrals of motion $H_1, H_2, H_3$ as follows:

$$H = \frac{cd^2}{(\frac{2e^{p_2}}{d} + \frac{x_1}{c})^2(-2ce^{p_2} - dx_1 + cdx_2)}, \tag{19}$$

$$H_1 = -\frac{2e^{p_2}}{d} - \frac{x_1}{c}, \tag{20}$$

$$H_2 = \frac{(-\frac{2e^{p_2}}{d} - \frac{x_1}{c})(-2ce^{p_2} - dx_1 + cdx_2)}{cd^2}, \tag{21}$$

$$H_3 = -\frac{(\frac{2e^{p_2}}{d} + \frac{x_1}{c})(-2ce^{p_2} - dx_1 + cdx_2)\log(|-2ce^{p_2} - dx_1 + cdx_2|)}{2cd^2}. \tag{22}$$

Let that $X^i$ of the vector field $X$ will be second order polynomials in position $x_1, x_2$. the Poisson Second bi-vector will be as follows:

$$P' = \begin{bmatrix} 0 & -2c\, a_4 & c\, a_4 - 2c(a_2 + 2a_3 x_2) & -d\, e^{-p_2}(b_1 + 2b_2 x_2) \\ * & 0 & 0 & -a_4 d\, e^{-p_2} \\ * & * & 0 & -d\, e^{-p_2}(a_2 + 2a_3 x_2) \\ * & * & * & 0 \end{bmatrix}, \tag{23}$$

where $a_i$s and $b_i$s are arbitrary real constants.
In this case for the Poisson bivector $P'$ (23), control matrix $F$ reads as

$$F = \begin{bmatrix} \dfrac{\alpha_1}{(c(2ce^{p_2} + dx_1)(2ce^{p_2} + dx_1 - cdx_2)))} & \dfrac{\alpha_2}{((2ce^{p_2} + dx_1)^4(2ce^{p_2} + dx_1 - cdx_2)^3)} \\ \dfrac{\alpha_3}{c^6 d^7} & \dfrac{\alpha_4}{(c(2ce^{p_2} + dx_1))} \end{bmatrix}, \tag{24}$$

where

$$\alpha_1 = -((a_4 c(2ce^{p_2} + dx_1)(2ce^{p_2} + dx_1 - cdx_2) + b_1(6d^2 x_1^2 + cdx_1(24e^{p_2} - 7dx_2) + \\ 2c^2(12e^{2p_2} - 7de_2^p x_2 + d^2 x_2^2)) + 2b_2 x_2(6d^2 x_1^2 + cdx_1(24e^{p_2} - 7dx_2) + \\ 2c^2(12e^{2p_2} - 7de^{p_2} x_2 + d^2 x_2^2))),$$

$$\alpha_2 = (c^4 d^7(b1 + 2b_2 x_2)(6ce^{p_2} + 3dx_1 - 2cdx_2)^2),$$

$$\alpha_3 = ((2ce^{p_2} + dx_1)^2 (b_1 + 2b_2 x_2)(4ce^{p_2} + 2dx_1 - cdx_2)^2(-2ce^{p_2} - dx_1 + cdx_2)),$$

and

$$\alpha_4 = (-a_4 c(2ce^{p_2} + dx_1) + b_1(4ce^{p_2} + 2dx_1 - cdx_2) + 2b_2 x_2(4ce^{p_2} + 2dx_1 - cdx_2) + \\ \frac{((b_1 + 2b_2 x_2)(4ce^{p_2} + 2dx_1 - cdx_2)^2)}{(2ce^{p_2} + dx_1 - cdx_2)}.$$

**Lie group $A_{4,3}$:**
From [5], we have the following forms for the non-degenerate Poisson structure on $A_{4,3}$:

$$\{x_1, x_2\} = c\, j_2\, e^{-p_2}, \quad \{x_1, p_1\} = d\, e^{-p_2}, \quad \{x_1, p_2\} = h\, e^{-p_2}, \quad \{x_2, p_1\} = f, \tag{25}$$

where c, d, h and f are arbitrary real constants, where in matrix form as follows:

$$P = \begin{bmatrix} 0 & c\,p_2\,e^{-p_2} & d\,e^{-p_2} & h\,e^{-p_2} \\ -c\,p_2\,e^{-p_2} & 0 & f & 0 \\ -d\,e^{-p_2} & -f & 0 & 0 \\ -h\,e^{-p_2} & 0 & 0 & 0 \end{bmatrix}. \tag{26}$$

On this Lie group, the Hamiltonian H and integrals of motion $H_1, H_2, H_3$ as follows:

$$H = \frac{e^{-p_2}p_1}{2d^2fh}(ch\,p_1^2 - 2d(fe^{p_2}x_1 - d\,x_2 + c\,p_1p_2)), \tag{27}$$

$$H_1 = -\frac{p_1}{d}, \tag{28}$$

$$H_2 = \frac{p_1}{d}\left(-\frac{x_1}{h} + \frac{de^{-p_2}x_2}{fh} + \frac{ce^{-p_2}p_1^2}{2fd} - \frac{ce^{-p_2}p_1p_2}{fh}\right), \tag{29}$$

$$H_3 = \frac{p_1}{d}\left(\frac{x_1}{h} - \frac{de^{-p_2}x_2}{fh} - \frac{ce^{-p_2}p_1^2}{2fd} + \frac{ce^{-p_2}p_1p_2}{fh}\right)\left(Ln\left(\left|\frac{x_1}{h} - \frac{de^{-p_2}x_2}{fh} - \frac{ce^{-p_2}p_1^2}{2fd} + \frac{ce^{-p_2}p_1p_2}{fh}\right|\right)\right). \tag{30}$$

Let that $X^i$ of the vector field $X$ will be second order polynomials in momenta $p_1, p_2$. the Poisson Second bi-vector will be as follows:

$$P' = \begin{bmatrix} 0 & \alpha & 0 & -e^{-p_2}(b_2d + b_1h) \\ * & 0 & 0 & -b_2f \\ * & * & 0 & 0 \\ * & * & * & 0 \end{bmatrix}, \tag{31}$$

where $a_i$s and $b_i$s are arbitrary real constants and

$$\alpha = -e^{-p_2}(a_1d + a_5h - b_2cp_1 + 2a_2dp_1 - 2b_3e^{p_2}fp_1 + a_3hp_1$$
$$+ b_1c(-1 + p_2) + a_3dp_2 - b_4e^{p_2}fp_2 + 2a_4hp_2 + b_2e^{p_2}fx_1).$$

Control matrix for the Poisson bivector $P'(31)$ is as follows:

$$F = \begin{bmatrix} -b_1 & \frac{\alpha}{2dfh} \\ 0 & 0 \end{bmatrix}, \tag{32}$$

where

$$\alpha = e^{-p_2}(p_1(2a_1d^2 + 2a_5dh - 4b_2cdp_1 + 4a_2d^2p_1 - 4b_3de^{p_2}fp_1 + 2a_3dhp_1 - b_2chp_1^2 +$$
$$2a_3d^2p_2 - 2b_4de^{p_2}fp_2 + 4a_4dhp_2 + 2b_2cdp_1p_2 + 2b_2de^{p_2}fx_1 - 2b_2d^2x_2) +$$
$$b_1(cp_1(-2d - 3hp_1 + 4dp_2) + 2d(e^{p_2}fx_1 - dx_2))).$$

**Lie group $A_{4,6}^{a,0}$:**
For this Lie group we have the following non-degenerate Poisson structure

$$\{x_1, p_2\} = d\,e^{-ap_2}, \quad \{x_2, p_1\} = c, \tag{33}$$

where a, c and d are arbitrary real constants. and can be represented in matrix form as follows:

$$P = \begin{bmatrix} 0 & 0 & 0 & d\,e^{-ap_2} \\ 0 & 0 & c & 0 \\ 0 & -c & 0 & 0 \\ -d\,e^{-ap_2} & 0 & 0 & 0 \end{bmatrix}. \tag{34}$$

On this Lie group, the Hamiltonian H and integrals of motion $H_1, H_2, H_3$ as follows:

$$H = \frac{e^{-\frac{(2e^{2ap_2}x_1)}{d}x_2^2}}{c^2} , \tag{35}$$

$$H_1 = \frac{x_2}{c} , \tag{36}$$

$$H_2 = \frac{e^{-\frac{(e^{2ap_2}x_1)}{d}} x_2 \cos(\frac{(e^{2ap_2}x_1)}{ad})}{c} , \tag{37}$$

$$H_3 = \frac{e^{-\frac{(e^{2ap_2}x_1)}{d}} x_2 \sin(\frac{(e^{2ap_2}x_1)}{ad})}{c} . \tag{38}$$

Let that $X^i$ of the vector field $X$ will be second order polynomials in momenta $p_1, p_2$. the Poisson Second bi-vector will be as follows:

$$P' = \begin{bmatrix} 0 & 0 & -d\,e^{-ap_2}(b_1 + 2b_2\,p_2) & -de^{-ap_2}(a_1 + 2a_2p_2 + a(a_3 + p_2(a_1 + a_2p_2))) \\ * & 0 & 0 & 0 \\ * & * & 0 & 0 \\ * & * & * & 0 \end{bmatrix} , \tag{39}$$

where $a_i$s and $b_i$s are arbitrary real constants.
Control matrix for the Poisson bivector $P'$(39) is as follows:

$$F = \begin{bmatrix} -aa_3 - a_1 - aa_1p_2 - 2a_2p_2 - aa_2p_2^2 & \frac{2}{c^2}e^{-\frac{2e^{2ap_2}x_1}{d}}x_2(aa_3c + a_1c + aa_1cp_2 + 2a_2cp_2 + aa_2cp_2^2 + b_1e^{ap_2}x_2 + 2b_2e^{ap_2}p_2x_2) \\ 0 & 0 \end{bmatrix}$$
(40)

**Lie group $A_{4,7}$:**
The non-degenerate Poisson structure for this Lie group has the following form [5]:

$$\{x_1, p_1\} = -2cp_1e^{-2p_2}, \quad \{x_1, p_2\} = ce^{-2p_2}, \quad \{x_2, p_1\} = 2ce^{-2p_2} , \tag{41}$$

where c is the arbitrary real constant.
On this Lie group, the Hamiltonian H and integral of motion $H_1$ as follows:

$$H = \frac{p_1(-1 - e^{2p_2} + e^{4p_2}x_1 + e^{4p_2}x_2p_1)}{2c} , \tag{42}$$

$$H_1 = -p_1 . \tag{43}$$

Let that $X^i$ of the vector field $X$ will be second order polynomials in momenta $p_1, p_2$. the Poisson Second bi-vector will be as follows:

$$P' = \begin{bmatrix} 0 & \alpha & 4ce^{-2p_2}p_1(c_3 + c_1p_1 + c_2p_1^2) & 2ce^{-2p_2}p_1(c_1 + 2c_2p_1) - 2ce^{-2p_2}(c_3 + c_1p_1 + c_2p_1^2) \\ * & 0 & -4ce^{-2p_2}(c_3 + c_1p_1 + c_2p_1^2) & -2ce^{-2p_2}(c_1 + 2c_2p_1) \\ * & * & 0 & 0 \\ * & * & * & 0 \end{bmatrix} , \tag{44}$$

where $a_i$s, $b_i$s and $c_i$s are arbitrary real constants and

$$\alpha = 2ce^{-2p_2}(a_1 + 2a_2p_1 + a_3p_2) + 2ce^{-2p_2}p_1(b_1 + 2b_2p_1 + b_3p_2) - ce^{-2p_2}(b_4 + b_3p_1 + 2b_5p_2) .$$

Control matrix for the Poisson bivector $P'$ (44) is as follows:

$$F = \begin{bmatrix} -2(c_3 + p_1(c_1 + c_2p_1)) & \frac{\beta}{4c} \\ 0 & -2(c_3 + p_1(c_1 + c_2p_1)) \end{bmatrix} , \tag{45}$$

where

$$\beta = e^{2p_2}p_1(2a_1\ e^{2p_2} - b_4\ e^{2p_2} - 8c_2p_1 + 4a_2\ e^{2p_2}p_1 - b_3\ e^{2p_2}p_1 + 2b_1\ e^{2p_2}p_1 +$$
$$4b_2\ e^{2p_2}p_1^2 + 2a_3\ e^{2p_2}p_2 - 2b_5\ e^{2p_2}p_2 + 2b_3\ e^{2p_2}p_1\ p_2 + 16c_2\ e^{2p_2}p_1\ x_1 +$$
$$16c_2\ e^{2p_2}p_1^2\ x_2 + c_1(-4 + 8e^{2p_2}(x_1 + p_1\ x_2))) \ .$$

**Lie group $A_{4,9}^1$:**
For this Lie group the non-degenerate Poisson structure has the following form [5]:

$$\{x_1, p_1\} = 2cp_1 e^{-2p_2}, \quad \{x_1, p_2\} = -c\ e^{-2p_2}, \quad \{x_2, p_1\} = -2c\ e^{-2p_2}\ , \tag{46}$$

where c is the arbitrary real constant.
On this Lie group, the Hamiltonian H and integral of motion $H_1$ as follows:

$$H = -p_1\ , \tag{47}$$
$$H_1 = -e^{-2p_2}\ . \tag{48}$$

The Poisson Second bi-vector will be as follows:

$$P' = \begin{bmatrix} 0 & \alpha & -4ce^{-2p_2}p_1(c_3 + c_1p_1) & -2c_1ce^{-2p_2}p_1 + 2ce^{-2p_2}(c_3 + c_1p_1) \\ * & 0 & 4ce^{-2p_2}(c_3 + c_1p_1) & 2c_1ce^{-2p_2} \\ * & * & 0 & 0 \\ * & * & * & 0 \end{bmatrix}, \tag{49}$$

where $a_i$s, $b_i$s and $c_i$s are arbitrary real constants and

$$\alpha = -2ce^{-2p_2}(a_1 + 2a_2p_1 + a_3p_2) - 2ce^{-2p_2}p_1(b_1 + 2b_2p_1 + b_3p_2) + ce^{-2p_2}(b_4 + b_3p_1 + 2b_5p_2)\ .$$

Control matrix for the Poisson bivector $P'$ (54) is as follows:

$$F = \begin{bmatrix} -2(c_3 + c_1p_1) & 0 \\ 2c_1e^{-2p_2} & 2(c_3 + c_1p_1) \end{bmatrix}\ . \tag{50}$$

**Lie group $A_{4,12}$:**
Finally, for this Lie group we have the following non-degenerate Poisson structure [5]:

$$\begin{aligned}\{x_1, p_1\} &= -ce^{-p_1}(a\cos(p_2) + b\sin(p_2)), \\ \{x_1, p_2\} &= ce^{-p_1}(-b\cos(p_2) + a\sin(p_2)), \\ \{x_2, p_1\} &= ce^{-p_1}(b\cos(p_2) - a\sin(p_2)), \\ \{x_2, p_2\} &= -ce^{-p_1}(a\cos(p_2) + b\sin(p_2))\ ,\end{aligned} \tag{51}$$

where $c = \dfrac{1}{a^2 + b^2}$ and a, b are arbitrary real constants.
On this Lie group, the Hamiltonian H and integral of motion $H_1$ as follows:

$$H = -e^{-2p_2}\ , \tag{52}$$
$$H_1 = -e^{-p_1}\ . \tag{53}$$

The Poisson Second bi-vector will be as follows:

$$P' = \begin{bmatrix} 0 & \alpha & c(c_1e^{-p_1}(-b\cos(p_2) + a\sin(p_2))) & c(c_1e^{-p_1}(a\cos(p_2) + b\sin(p_2))) \\ * & 0 & -c(c_1e^{-p_1}(a\cos(p_2) + b\sin(p_2))) & c_1e^{-p_1}(-b\cos(p_2) + a\sin(p_2)) \\ * & * & 0 & 0 \\ * & * & * & 0 \end{bmatrix}, \tag{54}$$

where $a_i$s, $b_i$s and $c_i$s are arbitrary real constants and

$$\alpha = c(e^{-p_1}((a(-a_4 + b_1 - a_3 p_1 + 2b_2 p_1 - 2a_5 p_2 + b_3 p_2) + b(a_1 + b_4 + 2a_2 p_1 + b_3 p_1 + a_3 p_2 + 2b_5 p_2))\cos(p_2) -$$
$$(b(a_4 - b_1 + a_3 p_1 - 2b_2 p_1 + 2a_5 p_2 - b_3 p_2) + a(a_1 + b_4 + 2a_2 p_1 + b_3 p_1 + a_3 p_2 + 2b_5 p_2))\sin(p_2))) \ .$$

Control matrix for the Poisson bivector $P'$ (54) is as follows:

$$F = \begin{bmatrix} 0 & -2c_1 e^{p_1 - 2p_2} \\ \frac{1}{2} c_1 e^{p_1 - 2p_2} & 0 \end{bmatrix} \ . \tag{55}$$

# 4  Conclusion

As a main result, we find the bi-Hamiltonian structures for the list of symplectic four dimensional real Lie groups which are classified by G.Ovando in ([6]). Also we calculate the control matrices for their integrals to show that these integrals are bi-involutive.